\documentclass[aps,preprintnumbers]{revtex4}
\usepackage{amsfonts,amssymb,amsmath}            
\usepackage[]{graphics,graphicx,epsfig}            
\usepackage{amsthm}
\usepackage{subfigure}
\def\identity{\leavevmode\hbox{\small1\kern-3.8pt\normalsize1}}

\newcommand{\ket}[1]{\left | \, #1 \right\rangle}

\newcommand{\half}{\mbox{$\textstyle \frac{1}{2}$}}

\begin{document}

\title{Geometric Effects and Computation in Spin Networks}
\date{\today}

\author{Alastair \surname{Kay}}
\affiliation{Centre for Quantum
Computation,
             DAMTP,
             Centre for Mathematical Sciences,
             University of Cambridge,
             Wilberforce Road,
             Cambridge CB3 0WA, UK}
\author{Marie \surname{Ericsson}}
\affiliation{Centre for Quantum
Computation,
             DAMTP,
             Centre for Mathematical Sciences,
             University of Cambridge,
             Wilberforce Road,
             Cambridge CB3 0WA, UK}

\begin{abstract}
When initially introduced, a Hamiltonian that realises perfect
transfer of a quantum state was found to be analogous to an
$x$-rotation of a large spin. In this paper we extend the analogy
further to demonstrate geometric effects by performing rotations on
the spin. Such effects can be used to determine properties of the
chain, such as its length, in a robust manner. Alternatively, they
can form the basis of a spin network quantum computer. We
demonstrate a universal set of gates in such a system by both
dynamical and geometrical means.
\end{abstract}

\maketitle

\section{Introduction}

Many proposed schemes for quantum computation rely on
nearest--neighbour couplings in order to generate a two qubit gate.
However, during a computation, it is necessary to interact distant
qubits. Typically, one would imagine a series of nearest neighbour
SWAP operations to move two qubits together so that they can
interact. Recently, the question of more efficient protocols has
been addressed. In particular, a large set of different spin chains
that permit perfect state transfer are now known \cite{Christandl,
Christandl:2004a, Kay:2004c, transfer_comment,bose:2004a,shi:2004}.
The first of these, defined for an $N$ qubit chain, was based on an
analogy with a spin $J=\half(N-1)$ particle rotating about the
$x$-axis from its $\ket{M_z=J}$ to $\ket{M_z=-J}$ state. In this
paper we will show how to extend the analogy to achieve rotations
about the $z$- and $y$-axes. We demonstrate a protocol that uses
these in a geometric manner to determine properties of the chain,
such as its length.

In parallel to the development of state transfer in single chains,
Burgarth and Bose have concentrated on developing a protocol that
allows perfect state transfer in pairs of chains with less stringent
manufacturing requirements. Their initial demonstration \cite{Bos04}
involved encoding the state to be transmitted across the inputs of
two identical chains, but has culminated in a demonstration that the
two chains need not be identical \cite{Bose:2005a}. If we apply
geometric or topological effects during transmission of a quantum
state along a chain, then the action of the gate separates out from
the visibility, i.e. the arrival probability of the state. Since the
protocol of Burgarth and Bose guarantees perfect visibility, we can
consider using this to realise a robust computation. We construct a
universal set of spin networks that act on a quantum state as it is
transmitted. These chains require no external interaction at all.
However, the Hadamard gate is not geometric or topological in
nature. We thus go on to develop a geometric Hadamard gate, by
allowing adiabatic manipulation of particular inter-qubit couplings.

\section{Geometric Effects and Metrology of a Spin Chain} \label{sec:geometric}

The perfect state transfer chain of Christandl et al.
\cite{Christandl, Kay:2004c,Christandl:2004a} has a Hamiltonian of
the form
\begin{equation}
H=\lambda\sum_{i=1}^{N-1}\frac{\sqrt{i(N-i)}}{4}(\sigma_x^i
\sigma_x^{i+1}+\sigma_y^i\sigma_y^{i+1}).
\label{eqn:ham}
\end{equation}
This Hamiltonian has the property that it preserves total spin i.e.
if, at time $t=0$, there is a single excitation in the system then,
in the absence of external interactions, there will always be a
single excitation present. In the first excitation subspace, using
basis states $\ket{n}$ to denote the presence of the excitation on
qubit $n$, the Hamiltonian is
$$
\frac{\lambda}{2}\left(
\begin{array}{ccccc}
  0 & \sqrt{N-1} & 0 & ... & 0 \\
  \sqrt{N-1} & 0 & \sqrt{2(N-2)} &... & 0\\
  0 & \sqrt{2(N-2)} & 0 & ... & 0 \\
  0 & 0 & \sqrt{3(N-3)} & ...& 0 \\
  \vdots &  \vdots &  \vdots & \ddots & \sqrt{N-1} \\
  0 & 0 & 0 & \sqrt{N-1} & 0\\
\end{array}
\right),
$$
which is just the rotation matrix $J_x$ for a particle of spin
$J=\half(N-1)$, scaled by a strength parameter $\lambda$. If the
single excitation is on qubit $n$, then this is equivalent in the
spin picture to being in the state $\ket{M_z=J+1-n}$. We can thus
view state transfer between opposite ends of the chain as a rotation
by an angle $\pi$ around the $x$-axis of the Bloch sphere.
Similarly, we can use the Bloch sphere picture to understand that if
we rotate by an angle of $\pi/2$, we will move from the state
$\ket{M_z=J}$ to $\ket{M_y=J}$. However, with only the Hamiltonian
$J_x$, the rotations that we can generate are limited. We will thus
introduce a magnetic field gradient over the spin chain. This is
equivalent to a rotation around the $z$-axis of the spin-$J$
particle, and takes the form
$$
\frac{\kappa}{2}\left(
\begin{array}{ccccc}
  N-1 & 0 & 0 & ... & 0 \\
  0 & N-3 & 0 &... & 0\\
  0 & 0 & N-5 & ... & 0 \\
  \vdots &  \vdots &  \vdots & \ddots & 0 \\
  0 & 0 & 0 & 0 & 1-N\\
\end{array}
\right)
$$
in the first excitation subspace, where $\kappa$ is a coupling
strength that we control. The implementation of such a gradient
depends on the details of the physical system that we use to
implement our spins. If we consider systems such as quantum dots
\cite{lambropoulos} that are equally spaced, then this field is just
a linear gradient, which can be implemented comparitively easily and
without local control of the system.

\begin{figure}
\begin{center}
\includegraphics[width=0.3\textwidth]{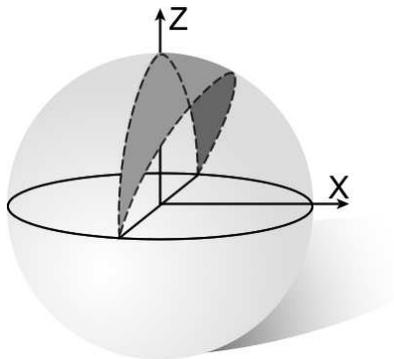}
\end{center}
\caption{By performing rotations around great circles, there are no
dynamical phases, making the result robust against some types of
error.} \label{fig:geometric}
\end{figure}

The situation that we envisage is that we have a chain with a fixed
Hamiltonian, eqn. (\ref{eqn:ham}), of known $\lambda$ and that we
can apply a gradient magnetic field. This is sufficient to be able
to implement a geometric rotation with the following protocol.
\begin{enumerate}
\item{Initialise the spin chain such that every spin is in the $\ket{0}$ state.}
\item{Place a single excitation at one end of the chain i.e. $\ket{M_z=J}$.}
\item{Wait for a time $\pi/(2\lambda)$, resulting in $\ket{M_y=J}$.}
\item{Switch on the magnetic gradient field at strength $\kappa$ for a
time $\pi/\sqrt{\lambda^2+\kappa^2}$, which rotates the state to $\ket{M_y=-J}$.}
\item{Switch off the magnetic field, and wait for a time $\pi/(2\lambda)$.}
\end{enumerate}
This protocol returns the single excitation to its starting point,
having followed great circles of the Bloch sphere, as depicted in
Fig. \ref{fig:geometric}. This means that no dynamic phase is
generated. However, a solid angle
$$
\Omega=2\tan^{-1}\left(\frac{\lambda}{\kappa}\right)
$$
has been carved out, and the geometric phase of the state is thus
\cite{geometric:a,geometric_bible}
$$
\gamma=\half(N-1)\Omega.
$$

Consider the case where a third party has manufactured a spin chain
for us, and assures us that the state transfer time,
$t_0=\pi/\lambda$, has a particular value. However, we do not know
what the length of the chain is. In this case, we can perform
interference experiments to determine the phase, and therefore the
number of spins on the chain. In the absence of errors, a modified
version of the semi-classical Fourier transform \cite{measuredFT},
combined with phase estimation, makes a particularly efficient
determination. To achieve this, we start without a magnetic field,
so that the state returns to the start of the chain with a phase of
$\pi(N-1)$ i.e. a phase factor of $\pm1$, determined by the parity
of the chain length. With the initial state
$(\ket{0}+\ket{1})/\sqrt{2}$ on the first qubit, this evolves to the
state $(\ket{0}-(-1)^N\ket{1})/\sqrt{2}$. After performing another
Hadamard on this qubit, it is in the state $\ket{0}$ or $\ket{1}$
depending on the parity of $N$. If we repeat this experiment,
generating half the phase (using the aforementioned protocol, Fig.
\ref{fig:geometric}), then the possible phase-factors are $i$, $-1$,
$-i$ and $1$. However, we know from the previous result that we will
either get one of the pair $i$ or $-i$ or one of the pair $-1$ or
$1$. If we know we're going to get the $i$ phases, then we
compensate for this with a $\sqrt{\sigma_z}$ gate before performing
the Hadamard. Measurement of the first spin then gives the second
least significant bit of the length of the chain. These steps can be
repeated (halving the solid angle, $\Omega$, each time) to determine
the next-least significant bit of $N$.

\begin{figure}
\begin{center}
\subfigure[Independent timing errors of up to $50\%$ for each of the 3 periods of evolution, averaged across 1000 runs.]{
\label{fig:timing}
\includegraphics[width=0.45\textwidth]{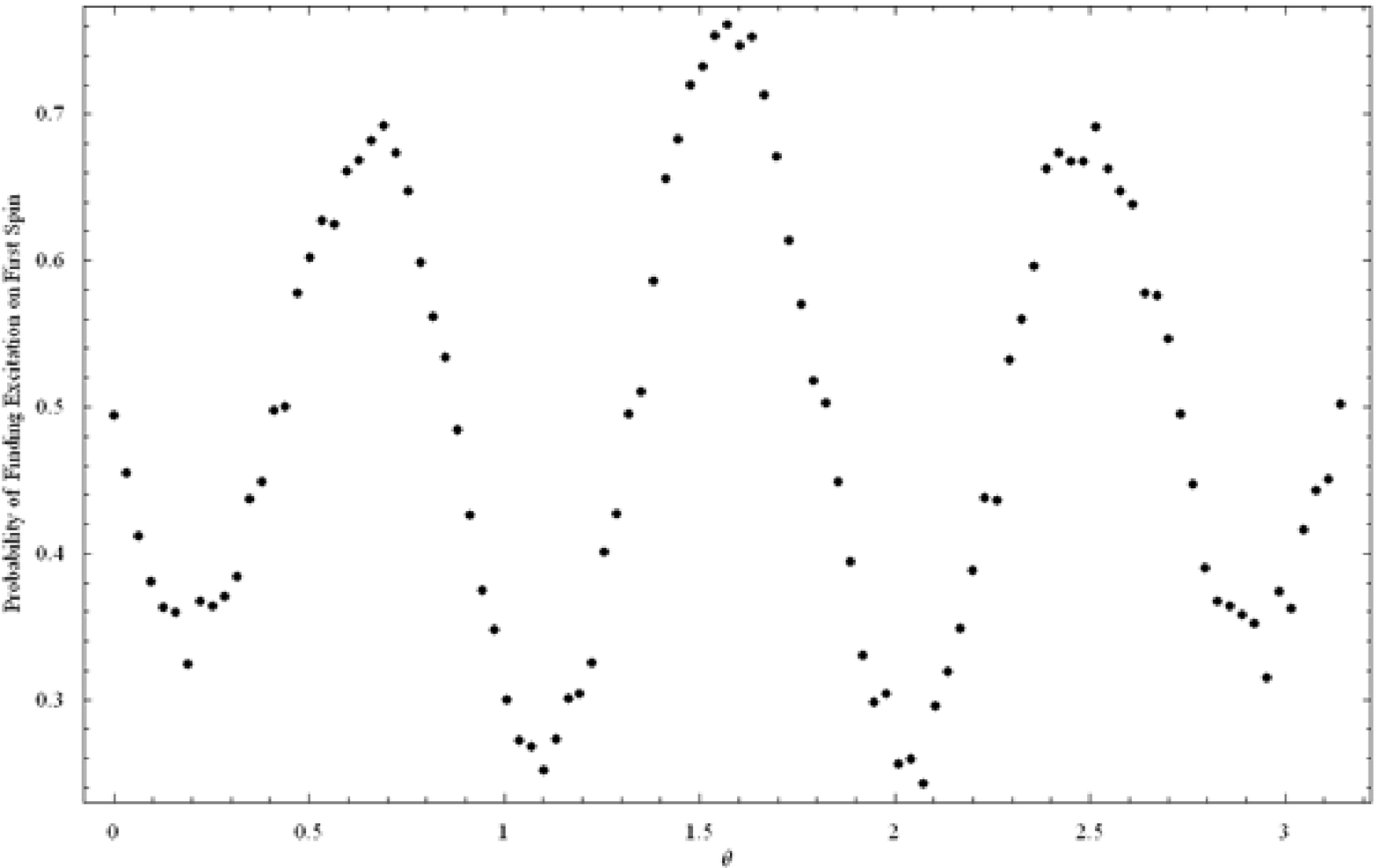}
}
\hspace{0.1cm}
\subfigure[Coupling errors of up to $80\%$ for each nearest-neighbour coupling in comparison to the ideal case. These errors are fixed for all experiments.]{
\label{fig:coupling}
\includegraphics[width=0.45\textwidth]{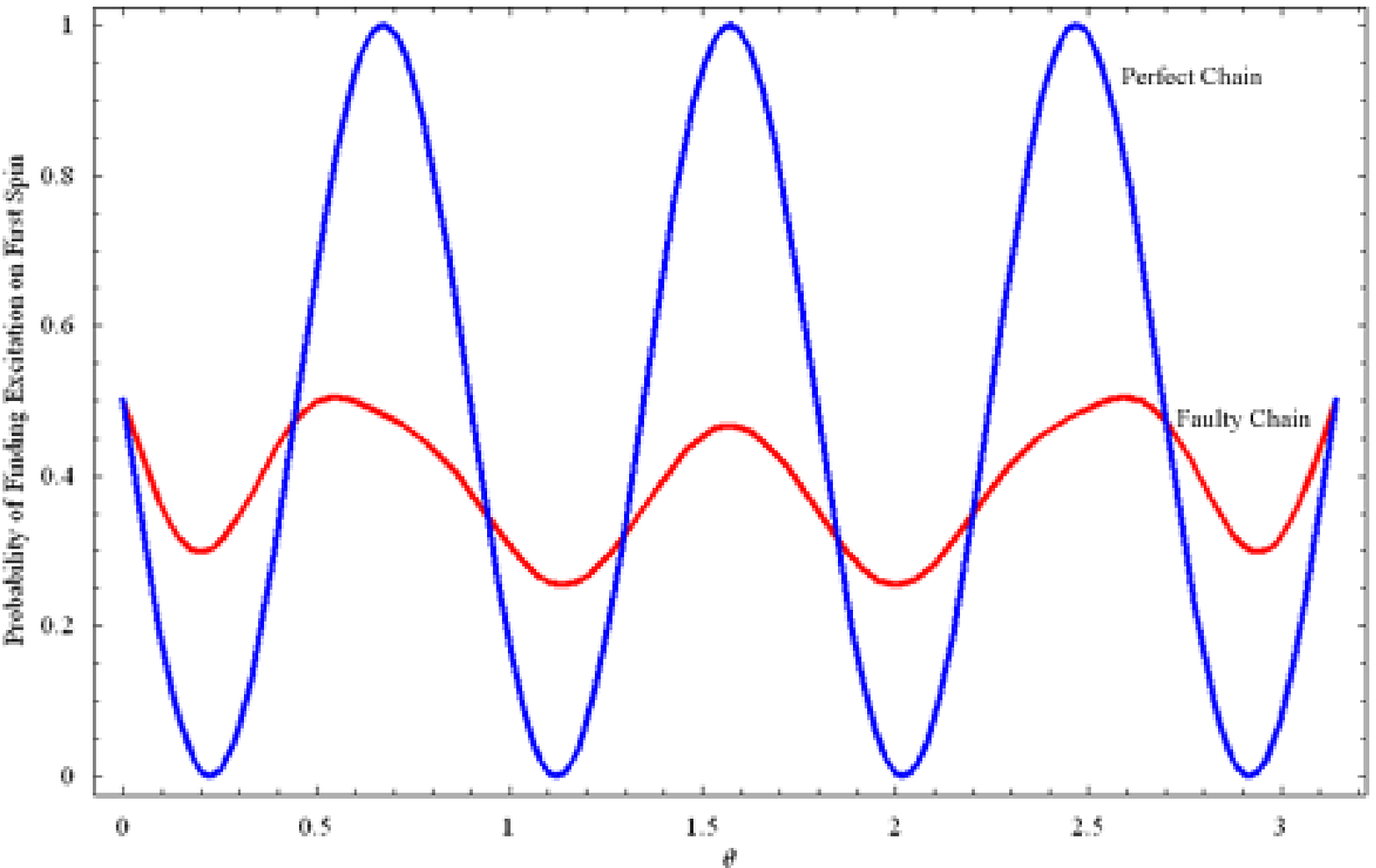}
}
\end{center}
\caption{The effect of errors of arrival probability of the state, as a function of magnetic field strength $\kappa=\cot(\theta)$ for a chain of 8 spins.}
\end{figure}

In the presence of errors, such as manufacturing errors or timing
errors, or in the presence of noise (interaction of the chain with
the environment), the semi-classical method fails because the
measurement results become probabilistic. Instead, there are two
ways that we can perform the experiment. Firstly, we might use the
more standard interferometric geometrical phase technique
\cite{wagh}, where we use the same magnetic field strength on every
experiment, and apply a different $z$-rotation, $e^{i\sigma_z\chi}$,
before applying the final Hadamard. The results that we plot, with
varying $\chi$, are well described by the function
$$
\half(1+\nu\cos(J\Omega-\chi))
$$
and the offset of the maximum, $\chi_0=J\Omega$, determines the
geometric phase, and thus the length of the chain. The quantity
$\nu$ is known as the visibility and encapsulates some information
about the errors that have occurred. However, it turns out that it
is better to set $\chi=0$ and plot the arrival probability with
varying $\Omega$ (i.e. the magnetic field strength). In this case,
we just estimate the frequency of the curve, and that tells us the
length of the chain. This method is superior because, not only is it
stable against timing errors (Fig. \ref{fig:timing}), but also
against errors in the coupling strengths (Fig. \ref{fig:coupling}),
which have no analogue in the spin picture.

When performing these experiments, we need to be aware that there
may be an offset to the magnetic field gradient i.e. the gradient is
the correct strength but the 0 of the field is not located on the
centre of the chain. This manifests itself as an additional,
dynamic, phase on the state. We can determine this shift using a
simple modification of the previous protocol. Once in the
$\ket{M_y=J}$ state, we turn on a magnetic gradient field of
strength $\kappa$ and wait for a time
$t=2\pi/\sqrt{\lambda^2+\kappa^2}$. Finally, we switch off the
magnetic field and wait for a time $3\pi/2\lambda$ so that our state
returns to its starting point. The geometric phase in this case is
$2\pi(N-1)$, thus leaving us simply with the dynamical phase due to
the offset of the magnetic field, which can be determined by a
series of interference experiments. Alternatively, this modified
protocol can be incorporated into the original one to remove the
effect of the offset directly.

\section{Spin Networks Based on Perfect Transfer}

So far, we have described how a spin chain with fixed couplings and
a controllable magnetic field gradient can yield a geometric phase
on a state. We will now demonstrate a powerful technique for
modifying any of the spin chains found in the literature
\cite{Christandl:2004a, transfer_comment, bose:2004a, shi:2004} to
give networks (i.e. no longer simple chains) that perform
non-trivial operations as a quantum state is transmitted along them.
The intuition behind how these modifications work stems from
understanding the projection method described in \cite{Kay:2004c},
which demonstrates that a hypercube, with spins associated with its
vertices, is equivalent to the perfect state transfer chain. When
such a projection is performed, there is a lot of freedom to choose
how the spins of the hypercube can be grouped, and it is this
freedom that we take advantage of.
\begin{figure}
\begin{center}
\subfigure[Spin chain that treats the state
$(\ket{01}+\ket{10})/\sqrt{2}$ like a single excitation on the
original perfect state transfer chain.]{ \label{fig:y}
\setlength{\unitlength}{0.75mm}
\begin{picture}(70,32)
\put(10,26){\circle*{2}} \put(10,6){\circle*{2}}
\put(20,16){\circle*{2}} \put(35,16){\circle*{2}}
\put(50,16){\circle*{2}} \put(65,16){\circle*{2}}
\put(10,16){\oval(4,30)}
 \put(15,6){\makebox(6,6){$\frac{J_1}{\sqrt{2}}$}}
 \put(15,19){\makebox(6,6){$\frac{J_1}{\sqrt{2}}$}}

 \put(-11,12){{\makebox(6,6){$\frac{1}{\sqrt{2}}(\ket{01}+\ket{10})$}}}

 \put(10,26){\line(1,-1){10}}
 \put(10,6){\line(1,1){10}}
 \put(20,16){\line(1,0){45}}

 \put(39,16){\makebox(6,6){$J_3\ldots$}}
 \put(26,16){\makebox(6,6){$J_2$}}
 \put(56,16){\makebox(6,6){$J_{N-1}$}}
\end{picture}
} \hspace{0.2cm} \subfigure[A perfect state transfer ring that could
have an electric charge or magnetic field threaded through it,
yielding a geometric or topological phase during transmission of a
state between A and B.]{ \label{fig:circle}
\setlength{\unitlength}{0.75mm}
\begin{picture}(80,30)
\put(15,22){\line(1,0){15}}
\put(15,2){\line(1,0){15}}
\put(45,2){\line(1,0){15}}
\put(45,22){\line(1,0){15}}
\put(5,12){\line(1,1){10}}
\put(5,12){\line(1,-1){10}}
\put(60,22){\line(1,-1){10}}
\put(60,2){\line(1,1){10}}

\put(5,12){\circle*{2}}
\put(15,22){\circle*{2}}
\put(15,2){\circle*{2}}
\put(30,22){\circle*{2}}
\put(30,2){\circle*{2}}
\put(45,22){\circle*{2}}
\put(45,2){\circle*{2}}
\put(60,22){\circle*{2}}
\put(60,2){\circle*{2}}
\put(70,12){\circle*{2}}

 \put(19,22){\makebox(6,6){$J_2$}}
 \put(35,19){\makebox(6,6){$...$}}
 \put(35,-1){\makebox(6,6){$...$}}
 \put(51,22){\makebox(6,6){$J_{N-2}$}}
 \put(19,2){\makebox(6,6){$J_2$}}
 \put(51,2){\makebox(6,6){$J_{N-2}$}}

 \put(-2,9){\makebox(6,6){A}}
 \put(72,9){\makebox(6,6){B}}

 \put(2,17){\makebox(6,6){$\frac{J_1}{\sqrt{2}}$}}
 \put(67,17){\makebox(6,6){$\frac{J_{N-1}}{\sqrt{2}}$}}
 \put(2,2){\makebox(6,6){$\frac{J_1}{\sqrt{2}}$}}
 \put(67,2){\makebox(6,6){$\frac{J_{N-1}}{\sqrt{2}}$}}
\end{picture}
}
\end{center}
\caption{Novel modifications of spin chains that result in state
transfer. $J_n$ represents an exchange coupling between the
indicated qubits, of strength $J_n=\lambda\sqrt{n(N-n)}/4$.}
\end{figure}
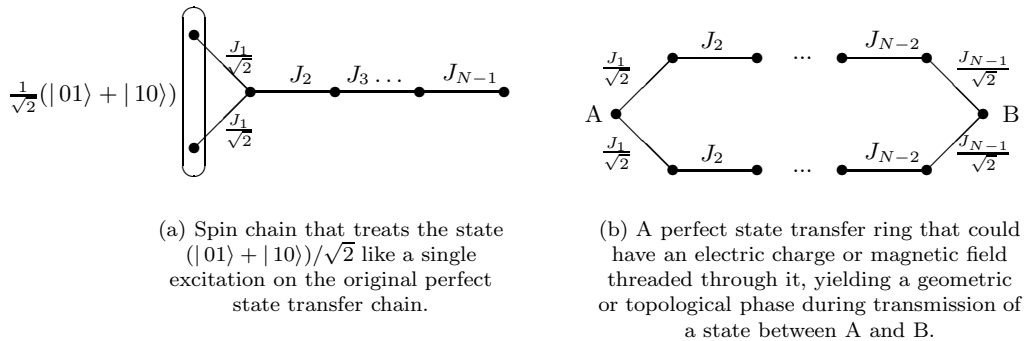

The two prime examples, from which all our modifications stem, are
shown in Figs. \ref{fig:y} and \ref{fig:circle}. Consider the
behaviour of the chain shown in Fig. \ref{fig:y}. Let us apply the
Hamiltonian of the chain to a state $(\ket{01}+\ket{10})/\sqrt{2}$,
defined across the left-most qubits. Both of these jump to the next
spin, $\ket{2}$, with a strength $J_1/\sqrt{2}$. So, we define
$$
\ket{\tilde 1}=(\ket{01}+\ket{10})/\sqrt{2},
$$
and find that $H\ket{\tilde 1}=J_1\ket{2}$, just as it does with the
original, unmodified, spin chain. If we now consider the action of
$$
H\ket{2}=\frac{J_1}{\sqrt{2}}(\ket{01}+\ket{10})+J_2\ket{3}=
J_1\ket{\tilde 1}+J_2\ket{3},
$$
we see, again, that it acts in exactly the same was as the original
chain, where the state $(\ket{01}+\ket{10})/\sqrt{2}$ is treated as
a single excitation. We can also verify that
$(\ket{01}-\ket{10})/\sqrt{2}$ is an eigenstate of the Hamiltonian,
with eigenvalue 0. This means that such a state remains localised on
those spins. Note that if we start with a quantum state on the
right-most spin, then in the perfect state transfer time, this
network acts as the optimal phase covariant cloner
\cite{spin_chain_clone}. The ring of Fig. \ref{fig:circle} also
behaves like a state transfer chain, where the excitations are
$(\ket{01}+\ket{10})/\sqrt{2}$ across the vertical pairs of qubits.
Note, however, that these projection methods only work for the first
excitation subspace. In the second excitation subspace there are
significant differences - a single chain has a fermionic character,
whereas the ring can allow two excitations in the pair of qubits.

\section{Quantum Computation by Transmission Through Spin Networks}

We have just seen that it is possible to split the original state
transfer chain into networks that still allow perfect state
transfer. Such modifications allow non-trivial unitary operations to
be applied to a state as it is transferred. Since the spin chains
are total spin preserving, the only operations that can be applied
to a single spin are phase gates, which are not sufficient for
universal quantum computation. Instead, we shall introduce the
dual-rail encoding
\begin{eqnarray}
\ket{0_L}&=&\ket{01}    \nonumber\\
\ket{1_L}&=&\ket{10}    \nonumber,
\end{eqnarray}
defined across the first qubit in each of two spin chains, which run
in parallel. When the state is transmitted, we will extract the
evolved state from the last qubit of each chain. Since these logical
states both contain a single excitation, any single--qubit state
contains a single excitation, and thus spin-preserving Hamiltonians
are able to perform one--qubit rotations.

This encoding is exactly that considered in \cite{Bos04,Bose:2005a}
for conclusive state transfer. In fact, any of the chains that we
construct will behave like two independent chains operating in
parallel, just with a basis change as the state moves along the
chain. As such, we can still use the techniques of Burgarth and Bose
to guarantee perfect visibility of the state, even in the presence
of manufacturing errors. All we have to do is perform some initial
experiments on the chains. Instead of plotting the arrival
probability with time for the two independent chains, and finding
times when the probabilities are equal, we plot the arrival
probability of a single spin across the two output qubits. The two
`chains' that we have to compare simply correspond to having
transmitted either the $\ket{0_L}$ or $\ket{1_L}$ states through the
network. Such testing before using the networks as part of a
computation also allows us to verify that the unitary operation is
of sufficiently high fidelity.

A universal set of gates for quantum computation can be created out
of an arbitrary $z$-rotation, a Hadamard and a two--qubit entangling
gate. It is worth noting that phase gates can be applied to the
dual-rail qubit by applying a phase gate to just one of the chains,
and other types of rotation can be effected by making swaps between
the two chains.

\subsection{Phase Gates}

Structures such as the ring of Fig. \ref{fig:circle} provide a
suitable topology for examining the Aharanov-Bohm
\cite{aharanov_bohm} or Aharanov-Casher effects
\cite{aharanov_casher}. The precise effect that we take advantage of
depends on the particular physical implementation of our spins. As a
concrete example, consider each spin being an empty quantum dot, and
we introduce a single electron to act as the excitation. This
situation has recently been discussed in the context of state
storage \cite{storage_rings}. In this realisation, the magnetic
gradient that we had previously is created with an electric field
gradient. By threading a magnetic field through the centre of the
ring, we realise the Aharanov-Bohm effect, which results in a phase
of $\phi$ if the electron moves around the ring. If there are $L$
links around the loop, then we can assume that the electron gains a
phase of $\delta=\phi/L$ as it jumps forwards, and $-\delta$ as it
jumps back again. Hence, in the first excitation subspace, the
Hamiltonian can be written as
$$
H=\cos(\delta)J_x+\sin(\delta)J_y.
$$
This just rotates the vector of our Hamiltonian by an angle $\delta$
about the $z$-axis. Hence, we still get perfect state transfer in
the time $\pi/\lambda$. As noted in \cite{Kay:2004c}, the state
gains an additional phase $\phi$ during transmission. This phase is
topological in nature, and thus makes a robust $z$-rotation of a
state during its transmission.

These rings are not limited to charged systems; we might
alternatively use rings of magnetically sensitive spins and take
advantage of the dual, Aharanov-Casher, effect \cite{spin_currents}.

\subsection{Hadamard Gate}
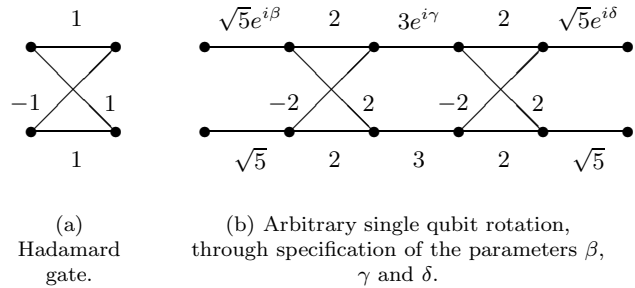
\begin{figure}
\begin{center}
\subfigure[Hadamard gate.]{
\label{fig:X}
\setlength{\unitlength}{0.75mm}
\begin{picture}(25,30)
\put(5,8){\line(1,0){15}}
\put(5,8){\circle*{2}}
\put(20,8){\circle*{2}}

 \put(10,0){\makebox(6,6){$1$}}

 \put(5,23){\line(1,0){15}}
\put(5,23){\circle*{2}}
\put(20,23){\circle*{2}}

 \put(10,25){\makebox(6,6){$1$}}

 \put(5,8){\line(1,1){15}}
 \put(5,23){\line(1,-1){15}}

 \put(16,10){\makebox(6,6){$1$}}
 \put(1,10){\makebox(6,6){$-1$}}
\end{picture}
} \subfigure[Arbitrary single qubit rotation, through specification
of the parameters $\beta$, $\gamma$ and $\delta$.]{
\label{fig:onequbit} \setlength{\unitlength}{0.75mm}
\begin{picture}(80,30)
\put(5,8){\line(1,0){75}}
\put(5,8){\circle*{2}}
\put(20,8){\circle*{2}}
\put(35,8){\circle*{2}}
\put(50,8){\circle*{2}}
\put(65,8){\circle*{2}}
\put(80,8){\circle*{2}}

 \put(10,0){\makebox(6,6){$\sqrt{5}$}}
 \put(25,0){\makebox(6,6){$2$}}
 \put(40,0){\makebox(6,6){$3$}}
 \put(55,0){\makebox(6,6){$2$}}
 \put(70,0){\makebox(6,6){$\sqrt{5}$}}

 \put(5,23){\line(1,0){75}}
\put(5,23){\circle*{2}}
\put(20,23){\circle*{2}}
\put(35,23){\circle*{2}}
\put(50,23){\circle*{2}}
\put(65,23){\circle*{2}}
\put(80,23){\circle*{2}}

 \put(10,25){\makebox(6,6){$\sqrt{5}e^{i\beta}$}}
 \put(25,25){\makebox(6,6){$2$}}
 \put(40,25){\makebox(6,6){$3e^{i\gamma}$}}
 \put(55,25){\makebox(6,6){$2$}}
 \put(70,25){\makebox(6,6){$\sqrt{5}e^{i\delta}$}}

 \put(20,8){\line(1,1){15}}
 \put(20,23){\line(1,-1){15}}
 \put(50,8){\line(1,1){15}}
 \put(50,23){\line(1,-1){15}}

 \put(31,10){\makebox(6,6){$2$}}
 \put(16,10){\makebox(6,6){$-2$}}
 \put(61,10){\makebox(6,6){$2$}}
 \put(46,10){\makebox(6,6){$-2$}}
\end{picture}
}
\end{center}
\caption{Two spin chain networks that are equivalent to perfect
state transfer chains, but perform non--trivial operations on
encoded qubits, $\ket{01}$ and $\ket{10}$, as they are transmitted
from left to right.}
\end{figure}
A Hadamard rotation is sufficient to convert $z$-rotations into any
arbitrary single--qubit gate. The creation of such a gate follows
from Fig. \ref{fig:y}, where a state of
$(\ket{01}+\ket{10})/\sqrt{2}$ is transferred as a single excitation
to the opposite end of a chain. Similarly, if we were to use an
altered form, where one of the $J_1/\sqrt{2}$ couplings is replaced
by $-J_1/\sqrt{2}$, then it would be the state
$(\ket{01}-\ket{10})/\sqrt{2}$ that gets transferred. Taking the
outputs of these two networks as the logical qubit, then we see that
a Hadamard rotation has been performed.
\begin{eqnarray}
\frac{1}{\sqrt{2}}(\ket{0_L}+\ket{1_L})&\rightarrow&\ket{0_L}   \nonumber\\
\frac{1}{\sqrt{2}}(\ket{0_L}-\ket{1_L})&\rightarrow&\ket{1_L}   \nonumber
\end{eqnarray}
This leads us to test the Hadamard gate as shown in Fig.
\ref{fig:X}, which is created by running these two structures in
parallel. We find that this acts as a unit which is capable of
replacing any section of parallel chains (acting with a strength
$\sqrt{2}$) and, as such, can be integrated to give circuits like
that of Fig. \ref{fig:onequbit} - an arbitrary one--qubit gate,
where the phases of $\beta$, $\gamma$ and $\delta$ result from
looping around magnetic fields (not shown) along the relevant
sections.

\subsection{Two--Qubit Gates}

We have demonstrated the ability to create an arbitrary
single--qubit rotation, acting on a dual rail qubit, by using two
parallel spin chains and interlinking them. We need no control after
initial manufacture if we assume perfect circuits. We would
therefore like to address the question of whether or not we can
create a complete circuit of spins for performing a quantum
algorithm, without the requirement of having to interact with the
circuit. To do this, we need to show how to integrate a two--qubit
gate, which would complete a universal set of gates.

One of our single qubit gates, the Hadamard, took the form of a unit
that could, by suitable scaling of the couplings, be used to replace
any sections of chain that run in parallel. We would like to find a
two--qubit gate that behaves in a similar way. So, assuming that we
have two qubits (i.e. four chains), is it possible to create an
entangling operation between them? The problem with connecting four
chains where the two pairs of chains have a single excitation each
is that we are no longer in the first excitation subspace and
perfect state transfer is not guaranteed in the second excitation
subspace.

Without a two--qubit gate that can be integrated into our networks,
we need to demonstrate some other form of two--qubit gate, and show
how to move states between it and the one--qubit gate networks. In
fact, the original state transfer chain has been shown to apply
controlled-phase gates between all states stored on the chain
\cite{bose:2004a,Jaksch:2004a}. This is because the chain inverts
the order of states during transmission. Since these states behave
like fermions, they gain a topological phase factor of -1 on
exchange. This can be used to create a suitable interaction between
our logical qubits, by placing one of the qubits from each pair on
the chain, and allowing state transfer to take place. This
generalises to creating multi-qubit entangled states by placing one
half of each logical qubit onto the chain.

\subsection{Joining Spin Chains}

With this form of two--qubit gate, we need to be able to join it
with other sections of chain, while minimising our interaction
with them. In particular, we are currently required to quickly
swap states between chains as they arrive (where `quick' is in
comparison to the transfer time $\pi/\lambda$). Previous works on
state transfer have implicitly assumed this for the purpose of
moving the state on and off the chain.

To do this, we introduce the idea of a switch, where we act on
specific qubits, decoupling them from neighbouring qubits. There are
two ways in which this can be implemented. Firstly, we can introduce
a large, local, detuning (i.e. a magnetic field)
\cite{BenjaminBose1}. However, this requires a finely tuned,
controllable, local field, which is technically difficult.

\begin{figure}
\begin{center}
\includegraphics[width=0.35\textwidth]{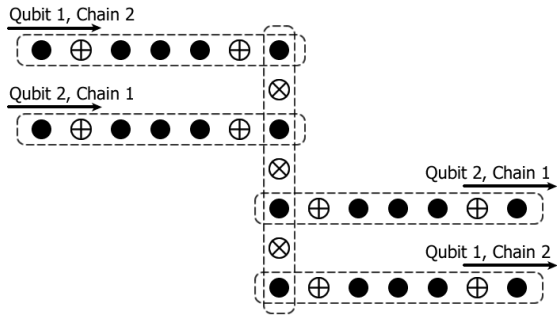}
\caption{The qubits $\otimes$ and $\oplus$ act as two different
types of switch. If they have an extra state, $\ket{2}$, which can
be populated by a global pulse, such as a laser, then when both
switch types are in this state, the state which we are computing is
trapped. Moving the one type of switch back to the $\ket{0}$ state
allows the computation to proceed in a fixed direction. The dashed
boxes indicated engineered spin chains which perform state
transfer.}\label{fig:switches}
\end{center}
\end{figure}

Instead, we consider the case where a spin has not only its
computational states, $\ket{0}$ and $\ket{1}$, but also an auxiliary
state, $\ket{2}$. By performing the transition
$\ket{0}\rightarrow\ket{2}$, this blocks the
$\sigma_x\sigma_x+\sigma_y\sigma_y$ coupling from causing the
excitation to jump to this spin. As demonstrated in Fig.
\ref{fig:switches}, two different types os switch are sufficient.
The switches $\otimes$ and $\oplus$ have different auxiliary states,
$\ket{2}$ and, as such, they can be addressed by different, global,
fields e.g. by different lasers. If the levels, $\ket{2}$, are at
the same energy as the computational states, then the accuracy of
this clocking method is only limited by the speed at which the
$\ket{0}\rightarrow\ket{2}$ transition can be performed in
comparison to the state transfer time. For an external field of
strength $\gamma$, the fidelity, $f\approx1-\alpha/\gamma^2$, since
this acts like a timing error on the state transfer process
\cite{Kay:2004c}. If the auxiliary levels are at energies
$E_\otimes$ and $E_\oplus$ above the levels of the computational
states, then spontaneous emission adds to the inaccuracy of the
scheme. This can be reduced by keeping the switching fields active
while the switches should be in the $\ket{2}$ state
\cite{benjamin:2004}.

\begin{figure}
\begin{center}
\includegraphics[width=0.4\textwidth]{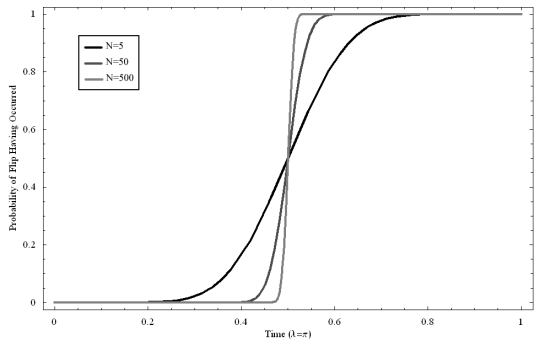}
\end{center}
\caption{As the length of the spin chain increases, we can achieve
an increasing sharp flip of the target spin at regular intervals.}
\label{fig:cnot}
\end{figure}

The required control over the two global fields can be reduced by
noting that the transition can be performed at fixed intervals of
$2\pi/\lambda$ if we create all our spin chains so that they have
the same state transfer time. In this situation, we can consider
internalising the switch (i.e. removing the global fields) by using
more spin chains. This is achieved by creating a form of
controlled-NOT gate, resulting from consideration of the Hamiltonian
$$
H=\sum_{i=1}^{N-1}K_i(\sigma_x^i\sigma_x^{i+1}+\sigma_y^i\sigma_y^{i+1})+
\eta(\sigma_z^N\otimes\sigma_x^{N+1}-\sigma_x^{N+1}).
$$
When we write out the Hamiltonian in a direct sum form, there is a
$2N\times 2N$ subspace that governs the evolution of the basis
states $\ket{n,0}$ and $\ket{n,1}$, where the $n$ represents the
location of a single excitation on the first $N$ qubits of the
chain, and the second component represents the state of the final,
target, qubit. This matrix is identical to the Hamiltonian of a
symmetric chain of length $2N$. We can therefore equate these
coefficients with those of a $2N$-spin perfect transfer chain,
$K_i=\sqrt{i(2N-i)}/4$ and $\eta=N/4$ such that in a time
$2\pi/\lambda$, we find that
$$\ket{1,0}\rightarrow\ket{1,1}.
$$
The $\ket{0,0}$ state remains unchanged, so this is a controlled-NOT
gate. The probability that the target qubit is in the $\ket{1}$
state is
$$
p=\sum_{n=N+1}^{2N}\sin^{2n-2}\left(\frac{\lambda t}{2}\right)
\cos^{4N-2n}\left(\frac{\lambda t}{2}\right)\binom{2N-1}{n-1}.
$$
Plotting this function (or writing as a hypergeometric function)
makes it clear that the result is an increasingly sharp flip of
the target qubit at times $(r+\half)\pi/\lambda$ (integer $r$) as
the length of the control chain increases (Fig. \ref{fig:cnot}).
Note, however, that the target of this operation is not a chain,
but a single qubit.

\begin{figure}
\begin{center}
\setlength{\unitlength}{0.75mm}
\begin{picture}(80,25)
\put(5,17){\line(1,0){75}}
\put(20,3){\line(0,1){15}}
\put(20,24){\vector(0,-1){5}}

\put(5,17){\circle*{2}}
\put(20,17){\circle*{2}}
\put(20,2){\circle*{2}}
\put(35,17){\circle*{2}}
\put(50,17){\circle*{2}}
\put(65,17){\circle*{2}}
\put(80,17){\circle*{2}}

 \put(10,18){\makebox(6,6){$J_1$}}
 \put(25,18){\makebox(6,6){$J_2$}}
 \put(40,18){\makebox(6,6){$J_3$}}
 \put(55,18){\makebox(6,6){$J_4$}}
 \put(70,18){\makebox(6,6){$J_5$}}
 \put(4,6){\makebox(6,6){$\gamma(\identity\!-\!\sigma_z)\!\otimes\!\sigma_z$}}
  \put(17,-4){\makebox(6,6){Target}}
   \put(17,24){\makebox(6,6){Switch}}
\end{picture}
\end{center}
\caption{The target of a controlled-NOT gate can conditionally
activate large Zeeman terms in the spin chain which cause reflection
rather than transmission. A single large Zeeman term, $\gamma$,
blocks transmission of states through the chain depending on the
lower, auxiliary qubit.} \label{fig:Zeeman_block}
\end{figure}
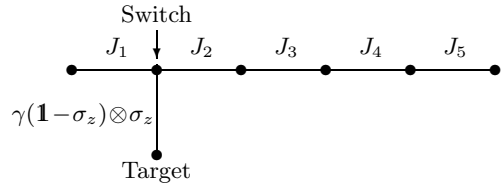

If we were able to implement this type of interaction (it has an
unphysical $\sigma_z\otimes\sigma_x$ term), we could choose to
implement the switch, where the $\sigma_x$ targets the switch qubit
on another chain, and acts between the $\ket{0}$ and $\ket{2}$
states. This results in a flip to the $\ket{2}$ state, which is
assumed to be stable. If we do not have a stable $\ket{2}$ state,
then the controlled-NOT can be used to target an auxiliary qubit,
which is connected as depicted in Fig. \ref{fig:Zeeman_block}. This
introduces a local detuning which is periodically switched on and
off, blocking the state transfer as required \cite{BenjaminBose1}.

\section{Non-Abelian Geometric Gates}

We have demonstrated how, in principle, we can create an entire
network of spins to perform a particular computation. However, we
have also seen that by introducing a limited form of control (a
magnetic field gradient), phase gates can be made geometric in
nature, yielding a robustness against a variety of errors. We will
now show how to create a Hadamard gate with geometric stability,
for which we invoke a non-abelian geometric phase.

\begin{figure}
\begin{center}
\setlength{\unitlength}{0.75mm}
\begin{picture}(130,40)

\put(5,8){\line(1,0){15}}
\put(40,8){\line(1,0){55}}
\put(115,8){\line(1,0){15}}
\put(5,8){\circle*{2}}
\put(20,8){\circle*{2}}
\put(40,8){\circle*{2}}
\put(55,8){\circle*{2}}
\put(80,8){\circle*{2}}
\put(95,8){\circle*{2}}
\put(115,8){\circle*{2}}
\put(130,8){\circle*{2}}
\put(27,5){\makebox(6,6){...}}
\put(102,5){\makebox(6,6){...}}

\put(5,33){\line(1,0){15}}
\put(40,33){\line(1,0){55}}
\put(115,33){\line(1,0){15}}
\put(5,33){\circle*{2}}
\put(20,33){\circle*{2}}
\put(40,33){\circle*{2}}
\put(55,33){\circle*{2}}
\put(80,33){\circle*{2}}
\put(95,33){\circle*{2}}
\put(115,33){\circle*{2}}
\put(130,33){\circle*{2}}
\put(27,30){\makebox(6,6){...}}
\put(102,30){\makebox(6,6){...}}

 \put(10,0){\makebox(6,6){$J_1$}}
 \put(3,10){\makebox(6,6){$\ket{1'}$}}
 \put(65,0){\makebox(6,6){$J_R\cos(\theta/2)$}}
 \put(120,0){\makebox(6,6){$J_{N-1}$}}
 \put(128,10){\makebox(6,6){$\ket{N'}$}}


 \put(10,35){\makebox(6,6){$J_1$}}
 \put(3,25){\makebox(6,6){$\ket{1}$}}
 \put(38,25){\makebox(6,6){$\ket{R\!-\!1}$}}
 \put(65,35){\makebox(6,6){$J_R\cos(\theta/2)$}}
 \put(120,35){\makebox(6,6){$J_{N-1}$}}
 \put(128,25){\makebox(6,6){$\ket{N}$}}

 \put(55,8){\line(1,1){25}}
 \put(55,33){\line(1,-1){25}}

 \put(86,15){\makebox(6,6){$-J_R\sin(\theta/2)e^{-i\phi}$}}
 \put(46,15){\makebox(6,6){$J_R\sin(\theta/2)e^{i\phi}$}}
\end{picture}
\end{center}
\caption{Two coupled perfect state transfer chains. Adiabatic
varying of the parameters $\theta$ and $\phi$ gives a non-abelian
geometric phase. The logical states of the qubit are defined by two
degenerate eigenvectors.} \label{fig:nonabelian}
\end{figure}
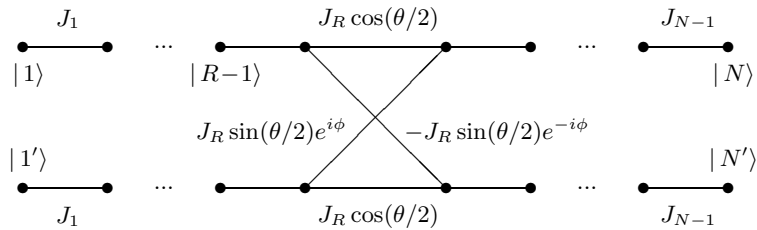

To realise a non-abelian geometric phase, we require spin networks
with degenerate eigenvalues. Consider, therefore, two perfect state
transfer chains, both of length $N$, where we retain the dual-rail
encoding. There are degenerate eigenvalues because everything is
repeated in comparison to a single chain. We can make some small
modification to these chains, linking them as shown in Fig.
\ref{fig:nonabelian}. This allows the introduction of two free
parameters, $\theta$ and $\phi$, that control the coupling. The
eigenvectors of maximum eigenvalue for this system are
\begin{eqnarray}
\ket{\psi_1}&=&\frac{1}{\sqrt{2^{N-1}}}\left(\sum_{n=1}^R\sqrt{\binom{N-1}
{n-1}}\ket{n}+\sum_{n=R+1}^N\sqrt{\binom{N-1}{n-1}}\left[\cos\left(\frac{\theta}{2}
\right)\ket{n}-\sin\left(\frac{\theta}{2}\right)
e^{-i\phi}\ket{n'}\right]\right)   \nonumber\\
\ket{\psi_2}&=&\frac{e^{-i\phi}}{\sqrt{2^{N-1}}}\left(\sum_{n=1}^R\sqrt{\binom{N-1}
{n-1}}\ket{n'}+\sum_{n=R+1}^N\sqrt{\binom{N-1}{n-1}}\left[\cos\left(\frac{\theta}{2}
\right)\ket{n'}+\sin\left(\frac{\theta}{2}\right)e^{i\phi}\ket{n}\right]\right), \nonumber
\end{eqnarray}
as can be readily verified. We can place a dual-rail encoded qubit
into these eigenstates by applying a magnetic gradient of strength
$\lambda$ for a time $\pi/(\sqrt{2}\lambda)$.

We will now consider a path in the parameter space $(\theta,\phi)$
that goes from
$(0,0)\rightarrow(\theta,0)\rightarrow(\theta,\phi)\rightarrow(0,\phi)$.
If we start our state in a degenerate eigenspace, then it remains in
this space when these parameters are varied adiabatically. By
differentiating the states $\ket{\psi_1}$ and $\ket{\psi_2}$ with
respect to $\theta$ and $\phi$ \cite{vlatko:2002}, we find that
$$
\frac{\partial}{\partial \zeta}\left(\begin{array}{c} a_1 \\ a_2
\end{array}\right)=A_\zeta\left(\begin{array}{c} a_1 \\ a_2
\end{array}\right)
$$
for $\zeta\in\{\theta,\phi\}$, where $a_i$ are the amplitudes of our
state in the two different eigenvectors. The matrices, $A_\zeta$,
are given by
\begin{eqnarray}
A_\theta&=&i\frac{P}{2}\sigma_y \nonumber\\
A_\phi&=&\frac{i}{2}(P\sin\theta\sigma_x+(1-P+P\cos\theta)\sigma_z) \nonumber
\end{eqnarray}
(up to an identity matrix, which only contributes a global phase) where
$$
P=\frac{2}{2^N}\sum_{n=R+1}^N\binom{N-1}{n-1}.
$$
Since both $A_\zeta$ are independent of $\zeta$, evaluating the
evolution just involves exponentiating the matrices. For
convenience, we define the variable $s$ such that $
A_\phi^2=-s^2\identity. $ We can therefore evaluate the whole
evolution as we vary the parameters
\begin{eqnarray}
U&=&e^{-i\frac{P\theta}{2}\sigma_y}e^{A_\phi\phi}e^{i\frac{P\theta}{2}
\sigma_y} \nonumber\\
&=&\identity\cos\left(s\phi\right)+\frac{i}{s}
e^{-iP\theta\sigma_y}\sin\left(s\phi\right)A_\phi. \nonumber
\end{eqnarray}
The Hadamard gate is defined as $H=(X+Z)/\sqrt{2}$, so we select $
\phi=\frac{\pi}{2s}, $ resulting in the evolution
$$
U=\frac{i}{s}\left((P\sin((P+1)\theta)+\sin(P\theta)(1-P))\sigma_x+
(\cos(P\theta)(1-P)+P\cos((P+1)\theta))\sigma_z\right).
$$
For a given value of $P$, we just have to solve for $\theta$. In
particular, for $R=N/2 \Rightarrow P=\half$, we find that
$\theta=\pi/4$ and $\phi=\pi/\cos(\pi/8)$. Similarly, if $R=1$, and
assuming a long chain, we find that $\theta\approx\pi/8$ and
$\phi\approx\pi$.

Finally, we should rotate back from the eigenstates of the chain to
being on the first (or last) spins of the chains by applying the
gradient magnetic field.

\section{Conclusions}

In this paper, we have demonstrated a number of widely applicable
techniques for modifying spin chains so that they can perform
non-trivial operations on encoded states. We have created a
universal set of both dynamic and geometric gates on these spin
chains. The geometric gates are particularly interesting because
these gates effectively make the statement ``if the state has
arrived, they have undergone the desired evolution". When coupled
with the conclusive state transfer protocol of Burgarth and Bose,
\cite{Bose:2005a}, this guarantees arrival of a correctly evolved
state, even in imperfect spin networks.

The geometric gates, while being much more robust, require a greater
degree of control over the system. The `dynamic' gates that we have
demonstrated require no external interaction after initial
manufacture (assuming they have a tolerable fidelity, which can be
tested), which result in the idea of fixed circuits of quantum
spins. Since they need no interaction (except for the input of
initial states, and the measurement of final states), we can, in
principle, completely isolate the system from the environment and
practically eliminate decoherence.

In addition, the techniques described here may be usefully
reinterpreted in many different contexts. For example, the
non-abelian geometric phase gate that we have demonstrated can be
translated into atomic systems, in a fundamentally different way to
the more standard `tripod' system \cite{unanyan,duan_tripod} that
has been extensively studied. It is, instead, like the less-studied
system introduced by Karle and Pachos \cite{Pachos:grass}.
Alternatively, the state transfer techniques may be of interest in
control theory. Here, the question of coherent population transfer
between the levels of an $N$-level atom  is considered. Cook and
Shore \cite{shore} originally demonstrated how a state can be
transferred from one level to another, using a single pulse, with a
scheme that is exactly equivalent to the original state transfer
chain. Coupling schemes such as that of Fig. \ref{fig:y}, and its
generalisation, would allow modification of that idea to transfer a
known superposition of two, or more, levels to another level of the
spin.

We thank Simon Benjamin, Artur Ekert and Daniel K.L. Oi for useful
discussions. ME acknowledge financial support from Swedish Research
council. AK is supported by UK EPSRC.

\end{document}